\documentclass[aps,twocolumn,showpacs,floatfix,amssymb]{revtex4}

\usepackage{amsmath}
\usepackage{amsfonts}
\usepackage{color}
\usepackage{graphicx}

\DeclareMathOperator{\sign}{sign}
\DeclareMathOperator{\tr}{tr}

\begin{document}
\title{Non-Fermi liquid criticality and super universality in the quantum Hall regime}
\author{A.\,M.\,M.\,Pruisken$^{1}$ and I.\,S.\,Burmistrov$^{1,2}$}

\affiliation{$^{1}$
Institute~for~Theoretical~Physics,~University~of~Amsterdam,~Valckenierstraat~65,~1018XE~Amsterdam,~The~Netherlands}
\affiliation{$^{2}$L.\,D.\,Landau Institute for Theoretical
Physics, Kosygina str. 2, 117940 Moscow, Russia }

\begin{abstract}
\noindent{We} report the results of a microscopic theory, based on
the topological concept of a $\theta$ vacuum, which show that the
Coulomb potential, unlike any finite ranged interaction potential,
renders the longstanding problem of the plateau transitions in the
quantum Hall regime non-Fermi liquid like. Our present results,
which are of outstanding significance for quantum phase
transitions in general and composite fermion ideas in particular,
provide a novel understanding of the critical exponent values that
have recently been (re)taken from a series of state-of-the-art
quantum Hall samples.
\end{abstract}

\pacs{71.10Pm, 72.10.-d, 73.43.-f}

\maketitle

\begin{table*}[tbp]
\begin{ruledtabular}
\caption{Coefficients of renormalization group functions, see
text. $A_2$ and $B_2$ for $0<c<1$ are yet unknown, $\gamma_E
\approx 0.577$ is the Euler constant and the analytic expressions
for $\mathcal{D} (c)$, $\mathcal{D}_\gamma (c)$ and $\mathcal{A}
\approx 1.64$ are given in Refs.~\cite{PruiskenBurmistrov2} and
\cite{UnifyII} respectively.}
\begin{tabular}{|c||c|c|c||c|c|c|}
 & $A_1$ & $A_2$ & $\mathcal{D}$ & $B_1$ & $B_2$ & $\mathcal{D}_\gamma$  \\
\hline $c = 0$&$0$~\cite{Brezin} & $1/2$~\cite{Brezin}&
$16\pi/e$~\cite{PruiskenBurmistrov1} & ${1}$~\cite{Pruisken84}&
$0$~\cite{Pruisken84}
& $8\pi/e$~\cite{PruiskenBurmistrov1} \\
$0\leq c \leq 1$ & ${2}[1 + \frac{1-c}{c} \ln
(1-c)]$~\cite{UnifyII,Italians} &$-$& $\mathcal{D}
(c)$~\cite{PruiskenBurmistrov2}
&${1}$~\cite{UnifyII,Italians}&$-$&$\mathcal{D}_\gamma(c)$
~\cite{PruiskenBurmistrov2}\\
$c = 1$ & ${2}$~\cite{Finkelstein} &
  $ {4 \mathcal{A}}$~\cite{UnifyII} &$16\pi
e^{1-4\gamma_E}$~\cite{PruiskenBurmistrov2} &${1}$
~\cite{Finkelstein} & $\pi^2/6 + 3$~\cite{UnifyII}& $8\pi
e^{1-4\gamma_E} /3$~\cite{PruiskenBurmistrov2}
\end{tabular}
\label{TZM1}
\end{ruledtabular}
\end{table*}

W. Li {\em et. al.}~\cite{W.} recently revisited the problem of
plateau transitions in the quantum Hall regime (QHR) by
investigating a series of specially grown samples with varying
disorder. Remarkably, both the experimental methodology and the
reported value of the universal exponent ($\kappa = 0.42$) are
exactly the same as those of the original papers by H.P. Wei {\em
et. al.}~\cite{Wei} more than 15 years ago. Like the original
experiments, W. Li {\em et. al.} put the all the emphasis on
samples with {\em short ranged potential fluctuations} relative to
the magnetic length. This fundamental aspect of the
theory~\cite{UnifyIII} has traditionally been
ignored~\cite{Schaijk} in the many experiments by many
others.~\cite{W.} Unfortunately, however, W. Li {\em
et.~al.}~\cite{W.} present their results based on Fermi-liquid
type of arguments that originally came along with the pioneering
work of H.P. Wei {\em et.~al.}~\cite{Wei} For example, the
observable exponent $\kappa = 0.42$ is generally recognized as the
ratio of two independent quantities, $\kappa = p/2\nu$, with $\nu$
denoting the correlation or localization length exponent and $p$ a
dynamical exponent that is determined by inelastic
processes.~\cite{Pruisken} W. Li {\em et. al.}~\cite{W.} identify
$\nu =2.3$ which is the free electron value known from numerical
work. The exponent $p$, on the other hand, which defines the
relation between the phase breaking length $L_\phi$ and
temperature $T$, popularly written as $L_\phi \propto T^{-p/D}$
with $D=2$ the spatial dimension,~\cite{Pruisken} has been taken
as a purely phenomenological quantity with a numerical value $p
\approx 2$.~\cite{W.}

In view of the dramatic progress made over the years, in the theory of
localization and interaction
effects,~\cite{UnifyI,UnifyII,UnifyIII} there actually exists no
reason to believe in Fermi liquid principles. Completely different
and novel insights have emerged which tell us that all the basic
aspects of the QHR reveal themselves as {\em super universal}
features of a single principle in quantum field theory, i.e. the
topological concept of an {\em instanton
vacuum}.~\cite{PruiskenBurmistrov1} The statement of super
universality makes it is easier and more natural to comprehend why
such basic phenomena as scaling and quantum criticality are
retained by the electron gas also in the presence of the
infinitely ranged Coulomb potential. Unlike Fermi liquid theory,
however, quantum criticality at $\theta=\pi$ does not necessarily
stay the same. Different applications of the {\em $\theta$ vacuum}
concept generally have different exponent values and, hence, fall
into different {\em universality
classes}.~\cite{PruiskenBurmistrov1}  {\em Super universality} is
important since it resolves many controversies ~\cite{LargeN} that
originally were encountered in QCD where the algebra is in many
ways the same. It has not been recognized until recently, for
example, that the $\theta$ vacuum concept generically displays
{\em massless chiral edge excitations}~\cite{UnifyIII} as well as
{\em robust topological quantum numbers}~\cite{LargeN} that
explain the {\em precision} and {\em observability} of the quantum
Hall plateaus. These very basic advances clearly indicate that the
QHR is an outstanding laboratory where the strong coupling
problems in quantum field theory can be explored and investigated
in great detail.

In this Letter we present the results of a detailed analysis on
topological excitations (instantons)~\cite{PruiskenBurmistrov2}
which unequivocally demonstrate that the experimental exponents
$\nu$ and $p$, rather than being disconnected pieces in heuristic
scenario's, are in fact universal quantities that emerge
simultaneously from a unifying microscopic theory.~\cite{UnifyI}
This theory describes the low energy dynamics of spin polarized
electrons in the presence of electron-electron interactions and
quenched ({\em short ranged}) disorder.~\cite{UnifyIII} It
encapsulates not only the different aspects of symmetry, quantum
criticality and super universality of the electron gas in the QHR
but also the conventional {\em mobility edge} problem in
$2+\epsilon$ dimensions~\cite{Finkelstein,Italians} which sets the
stage for our understanding of quantum phase transitions in
general.~\cite{UnifyII} For example, {\em exact} global symmetries
based on the electrodynamic $U(1)$ gauge invariance ({\em
$\mathcal F$ invariance}~\cite{UnifyI}) quite generally mark the
difference between a {\em Fermi liquid} and a {\em non-Fermi
liquid} universality class depending on whether the {\em range} of
the interactions is {\em finite} or {\em infinite}.~\cite{UnifyII}
The main objective of the present Letter is to elucidate the basic
principles of quantum transport theory and explain the
experimental results of W. Li {\em et. al.}~\cite{W.} based on a
novel non-Fermi liquid theory of the conductances.

The instanton vacuum representation of the spin polarized electron
gas involves unitary ({\em Grassmannian}) matrix field variables
${Q}_{mn}^{\alpha\beta}(\bf{r} )$ that obey $Q^2 (\bf{r} ) = {\bf
1}$. The integers $\alpha$, $\beta$ denote the {\em replica}
indices and $m$, $n$ correspond to the discrete set of Matsubara
frequencies $\omega_{n} = \pi T(2n+1)$. The effective action in
$D=2$ is given as
\begin{equation}
    S_\textrm{eff}[Q] =  S_\sigma [Q] +S_F [Q]
\end{equation}
\noindent{where} $S_{\sigma}$ represents the well know {\em free
electron} part~\cite{PruiskenBurmistrov1}
\begin{eqnarray}
    S_{\sigma} [Q] &=& -\frac{\sigma_{xx}}{8} ~\int d {\bf r} \tr
    \nabla_\mu Q \nabla_\mu Q
    + 2\pi i {\sigma_{xy}} ~ \mathcal{C} [Q] \label{freepart1}\\
    \mathcal{C} [Q] &=& \hspace{0.23cm}\frac{1}{16 \pi i}
    \int d {\bf r} \tr \epsilon_{\mu\nu}
    Q \nabla_\mu Q \nabla_\nu Q \label{freepart2}
\end{eqnarray}
with $\mathcal{C} [Q]$ denoting the topological charge and
$\sigma_{xx}$, $\sigma_{xy}$ the mean field parameters for the
longitudinal and Hall conductance respectively. The term $S_F$
contains the {\em singlet interaction} amplitude
$zc$~\cite{UnifyI} and has been proposed in different physical
settings before~\cite{Finkelstein,Italians}
\begin{equation}
    S_{F} = \pi z  T\hspace{-0.1cm}
    \int \hspace{-0.1cm} d \mathbf{r}\left[ c {\sum_{\alpha n}} \tr {I}_n^\alpha Q
    \tr {I}_{-n}^\alpha Q +4 \tr\eta  Q - 6 \tr\eta  \Lambda
    \right ] \label{SF}
\end{equation}
The meaning of the symbols is as follows. The theory in the range
$0 < c < 1 $ describes the electron gas with {\em finite} ranged
electron-electron interactions. The cases $c = 0$ and $c = 1$
describe {\em free} electrons and the problem with {\em
infinitely} ranged interaction potentials such as the Coulomb
potential respectively. The matrices $I_n$, $\eta$ and $\Lambda$
are all directly related to the electron dynamic gauge invariance
of the theory.~\cite{UnifyI} The $\eta$ and $\Lambda$ are diagonal
\begin{equation}
 \eta_{nm}^{\alpha\beta} = n {\bf 1}_{nm}^{\alpha\beta} ,\qquad
 \Lambda_{nm}^{\alpha\beta} = \sign (\omega_n )
 {\bf 1}_{nm}^{\alpha\beta}
\end{equation}
with $\eta$ standing for the {\em frequency} matrix and $\Lambda$
representing the {\em classical} value of $Q$ (a convenient and
frequently used representation is $Q = \mathcal{T}^{-1} \Lambda
\mathcal{T}$ with unitary $\mathcal{T}$). The $I_k$ are shifted
diagonal matrices
\begin{equation}
    [{I}_k^\gamma ]_{nm}^{\alpha\beta} = \delta_{\alpha\gamma}
    \delta_{\beta\gamma}\delta_{n,m+k}
\end{equation}
\noindent{which} span a $U(1)$ algebra $I_m^\alpha I_n^\beta =
\delta_{\alpha\beta} I_{m+n}^\alpha$. These are recognized as the
generators of electrodynamic gauge transformations in Matsubara
frequency space which can formally be represented by unitary
rotations $Q \rightarrow \mathcal{W}^{-1} Q
\mathcal{W}$.~\cite{UnifyI} One of the longstanding and notorious
complications of Eqs.~\eqref{freepart1}-\eqref{SF}, however, is
that the electrodynamic gauge invariance is preserved only if the
Grassmannian field variables $Q$ are taken as {\em infinite size}
matrices in frequency space. To ensure that the different limits
of the theory ({\em replica limit} along with {\em infinite}
frequency space) can generally be taken in a unique, cut-off
independent manner it is necessary to spell out the different
physical constraints that should be rigorously retained by
Eqs.~\eqref{freepart1}-\eqref{SF} for all values of the bare
parameters $\sigma_{xx}$, $\sigma_{xy}$, $c$ and $z$. The matter
has been investigated in considerable detail in a series of more
recent papers on abelian quantum Hall states based on an extended
version Eqs.~\eqref{freepart1}-\eqref{SF} that includes the Chern
Simons statistical gauge fields.~\cite{UnifyI,UnifyIII} The
results can be summarized as follows.

(i)~~The Coulomb interaction problem displays an {\em exact}
global symmetry, termed $\mathcal{F}$ {\em invariance}, which
means that the action with $c=1$ is invariant under global $U(1)$
gauge transformations. $\mathcal{F}$ {\em invariance} is broken by
finite ranged interactions $0 < c <1$ as well as the free electron
approximation $c=0$.~\cite{UnifyI}

(ii)~~In order for the theory to be consistent with the
macroscopic conservation laws (continuity equation) it is
imperative that the combination $z(1-c)$ in Eq.~\eqref{SF} does
not acquire radiative corrections, neither perturbatively nor in
the theory at a non-perturbative
level.~\cite{Generalization,UnifyII}

\begin{figure}[tbp]
\includegraphics[height=50mm]{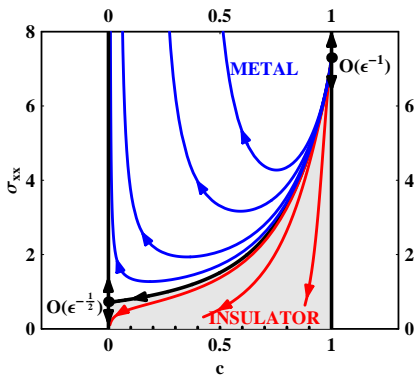}
\caption{Sketch of scaling in the $\sigma_{xx}$ versus $c$ plane
in $2+\epsilon$ dimensions. For illustration we have taken
$\epsilon=0.1$, see text.} \label{flowlines}
\end{figure}

\noindent{\em $\bullet$ $~D= 2+ \epsilon$.} To see these general
statements at work in explicit computations we discard the
topological charge in Eq.~\eqref{freepart1} for the moment and
present the results of the theory in $D = 2+\epsilon$ dimensions.
These have recently been extended to one order in $\epsilon$
higher than what was known previously.~\cite{UnifyII} The
renormalizations of the $\sigma_{xx}$ and $zc$ fields can be
expressed as follows (see Table~\ref{TZM1})
\begin{eqnarray}
    \beta_{\sigma}^0 (\sigma_{xx} ,c)&=& \frac{d \sigma_{xx}}{d\ln b}
    = \epsilon \sigma_{xx} - \frac{A_1 (c)}{\pi} - \frac{A_2 (c)}{\pi^2
    \sigma_{xx}} ~~~~~~
    \label{beta1}\\
    \gamma_{zc}^0 (\sigma_{xx} ,c) &=& \frac{d\ln zc}{d\ln b} = ~~~~
    - \frac{B_1 (c)}{\pi\sigma_{xx}} - \frac{B_2 (c)}{
    \pi^2\sigma_{xx}^{2}}.\label{beta2}
\end{eqnarray}
The dependence on $c$ is a peculiar feature of interaction terms
like $S_F$ which is quite unlike the conventional role played by
ordinary mass terms.~\cite{Brezin,Pruisken84} This dependence is,
in fact, the way in which electrodynamic gauge invariance
manifests itself in the theory of
Eqs.~\eqref{freepart1}-\eqref{SF}. For example, on the basis of
statement (ii) above one can express the equations for $c$ and $z$
in terms of $\gamma_{zc}^0$ as follows
\begin{eqnarray}
    \beta_c^0 (\sigma_{xx} ,c) &=& \frac{d c}{d\ln b}
    = c(1-c) \gamma_{zc}^0 (\sigma_{xx} ,c) ~~~~~~\label{beta3}
    \\
    \gamma_{z}^0 (\sigma_{xx} ,c) &=& \frac{d\ln z}{d\ln b} =
    ~~~~~~~~~c \gamma_{zc}^0 (\sigma_{xx} ,c) ~~\label{beta4}
\end{eqnarray}
which are in accordance with the results obtained in explicit
computations.~\cite{UnifyII} In Fig.~\ref{flowlines} we plot the
renormalization group flow lines in the $\sigma_{xx}$ versus $c$
plane. The results are reminiscent of those in the quantum Hall
regime, to be discussed below, in that the $\mathcal{F}$ {\em
non-invariant} theory $ 0<c < 1 $ generally lies in the domain of
attraction of the {\em Fermi liquid} line $c=0$ whereas the
$\mathcal{F}$ {\em invariant} theory $ c = 1 $ constitutes a {\em
non-Fermi liquid} universality class all by itself. There are two
different critical fixed points, one with $c^* =0$, $\sigma_{xx}^*
= \mathcal{O} (\epsilon^{-1/2})$ and one with $c^*=1$,
$\sigma_{xx}^* = \mathcal{O} (\epsilon^{-1})$, which separate a
weak coupling metallic state from a strong coupling insulating
phase. The two independent critical exponents $\nu$ and $p$
obtained as
\begin{equation}
 \nu^{-1} = \partial\beta_\sigma^*/\partial \sigma_{xx},
 \qquad p = D/(D+\gamma^*_{zc}) \label{NFL}
\end{equation}
have exactly the same meaning as those describing the quantum Hall
plateau transitions. The non-Fermi liquid values obtained from the
$c^* =1$ fixed point are given by $\nu^{-1} = \epsilon
+\mathcal{A}\epsilon^2$ and $p= 1+(\epsilon/8)+[(\pi^2+15)/12
-\mathcal{A}]\epsilon^2/16$. For finite range potentials $0< c <1$
the problem generally involves three independent scaling fields.
For example, let $\Delta\sigma \approx (\sigma_{xx} -
\sigma_{xx}^*)/\sigma_{xx}^* -c$ and $c$ be the small deviations
from the Fermi liquid fixed point then the scaling behavior of
conductivity $\sigma_{xx}^\prime$ is expressed as follows
\begin{equation}
 \sigma_{xx}^\prime = b^\epsilon \sigma_{xx}^\prime (
 b^{-1/\nu} \Delta\sigma; b^{D-D/p}c; b^{-D/p} z c T)
 \label{Scaling}
\end{equation}
which generalizes the free particle theory ($c =0$) where the $T$
dependence is strictly absent. Besides different exponent values
$\nu^{-1} = 2 \epsilon + 3\epsilon^2$ and $p= 1+\sqrt{2\epsilon}$
there are also important physical differences in fundamental
aspects such as the {\em multi fractal} singularity spectrum, the
specific heat of the electron gas etc. which usually are not being
probed in transport measurements.
~\cite{PruiskenBurmistrov1,UnifyII,PruiskenBurmistrov2}

%
%

\noindent{$\bullet$} {\em QHR.~} The metallic phases disappear all
together in $D=2$. This dramatic conflict with the QHR demands a
complete understanding of the topological piece in
Eq.~\eqref{freepart1}. To explore the fundamental significance of
the aforementioned {\em massless chiral edge excitations} we next
introduce a change of variables $Q=t^{-1} Q_0
t$.~\cite{PruiskenBurmistrov1} Here, the $Q_0$ represents an
arbitrary `{\em bulk}' matrix field configuration with the
classical value $Q_0 =\Lambda$ at the edge. The $t$ are recognized
as the `{\em edge}' sector of the theory describing the
fluctuations about these very special boundary conditions. The
distinctly different properties of the `{\em bulk}' and `{\em
edge}' sectors of the theory can generally be expressed in terms
of an effective action for the `{\em edge}'
\begin{eqnarray}
\exp S_\textrm{eff}^\prime [q] & = & \int_{\partial V} D[Q_0] \exp
S_\textrm{eff}[t^{-1} Q_0 t].\label{Sprime}
\end{eqnarray}
Here, $q = t^{-1} \Lambda t$ and the symbol ${\partial V}$ reminds
us of the fact that the theory is evaluated with fixed boundary
conditions $Q_0 = \Lambda$. The theory of Eq.~\eqref{Sprime}, as
it now stands, provides a complete description of the low energy
dynamics of the electron gas for all values of $\sigma_{xx}$,
$\sigma_{xy}$, $c$ and
$z$.~\cite{PruiskenBurmistrov1,PruiskenBurmistrov2} To show this
we elaborate on two extremely important limits of the problem.
First, there is the `naive' strong coupling limit of the theory
which physically corresponds the situation where the Fermi energy
is located in a Landau gap. In this case, all the bare parameters
$\sigma_{xx}$, $z$ and $c$ in Eqs.~\eqref{freepart1} and
\eqref{freepart2} are identically zero except $\sigma_{xy} = k$
where the integer $k$ denotes the number of completely filled
Landau levels. The action now contains the topological piece only
and can be written as~\cite{UnifyIII}
\begin{equation}
S_\textrm{eff} [t^{-1} Q_0 t ] \rightarrow \frac{k}{2} \oint
 dx \tr  {t}  \nabla_x {t}^{-1} \Lambda
 + 2\pi i~k~\mathcal{C}[Q_0]  \label{Landaugap1}
\end{equation}
where the integral is over the edge of the system. Since
$\mathcal{C}[Q_0]$ is by construction integer valued we conclude
that $S_\textrm{eff}^\prime [q] = S_\textrm{eff} [q]$, discarding
constants and phase factors that are immaterial. Surprisingly,
this one dimensional theory is exactly solvable and the spectrum
consists of {\em massless chiral edge excitations} which are
completely independent of the details such as the geometry of the
edge, the number of field components in the theory
etc.~\cite{UnifyIII} It is not difficult to see that
$S_\textrm{eff}^\prime [q]$ must in general be of exactly the same
form as Eq.~\eqref{Landaugap1} provided the bulk sector $Q_0$ in
Eq.~\eqref{Sprime} develops a {\em mass gap}.
Eq.~\eqref{Landaugap1} is therefore quite generally recognized as
the {\em fixed point action} of the quantum Hall state with the
integer $k$ now standing for the {\em robustly quantized Hall
conductance} rather than the filling fraction. The most important
task next, however, is to show whether and how this super
universal {\em strong coupling} feature of the $\theta$ vacuum can
be reconciled with the {\em weak coupling} results of
Eqs~\eqref{beta1}-\eqref{beta4}. This takes us to the second limit
of the theory where $S_\textrm{eff}^\prime [q]$ is formally
evaluated in terms of a series expansion in powers of both $T$ and
the derivatives acting on $q$. Of interest are the lowest order
terms which display exactly the same form as the original action
$S_\textrm{eff}[Q]$ except that the bare parameters in
Eqs~\eqref{freepart1} and \eqref{freepart2} are now replaced by
the {\em observable} quantities $\sigma_{xx}^\prime$,
$\sigma_{xy}^\prime$, $z^\prime$ and $c^\prime$ respectively.
~\cite{PruiskenBurmistrov1,PruiskenBurmistrov2,UnifyII}. The
expressions for $\sigma_{xx}^\prime$ and $\sigma_{xy}^\prime$,
obtained in this way, can quite generally be identified with the
linear response formulae for the conductances at $T=0$ which now
appear as a measure for the response of the bulk of the system to
infinitesimal changes in the boundary conditions. The $z^\prime$
and $z^\prime c^\prime$, on the other hand, are analogous to the
spontaneous magnetization in the classical Heisenberg
ferromagnet.~\cite{Brezin} The relation between the {\em
observable} and {\em bare} theories generally defines $\beta$ and
$\gamma$ functions according to ~\cite{PruiskenBurmistrov2}
\begin{eqnarray}
    \sigma_{xx}^\prime &=& \sigma_{xx}^0 + \int_{b_0}^{b^\prime} \frac{db}{b}
    \beta_\sigma ,~~~~
    \sigma_{xy}^\prime = \sigma_{xy}^0 + \int_{b_0}^{b^\prime} \frac{db}{b}
    \beta_\theta \\
    z^\prime c^\prime &=& z_0 c_0 + \int_{b_0}^{b^\prime} \frac{db}{b}
    zc
    \gamma_{zc} , ~~z^\prime = z_0 ~+ \int_{b_0}^{b^\prime} \frac{db}{b} z
    \gamma_{z} .
\end{eqnarray}
%
\begin{figure}[tbp]
\includegraphics[height=50mm]{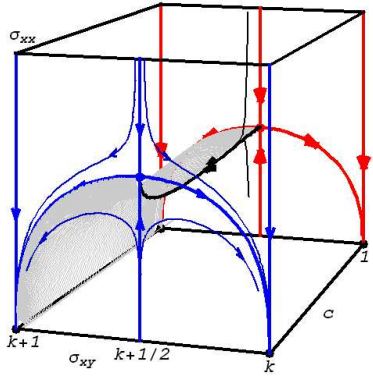}
\caption{Sketch of the renormalization group flow in the parameter
space $\sigma_{xx}$, $k \leq \sigma_{xy} \leq k+1$ and $0 \leq
c\leq 1$, see text.} \label{RGall}
\end{figure}
%
\noindent{A} complete quantum theory which includes instanton
contributions to the $\beta$ and $\gamma$ functions has been
developed only recently. The final results satisfy the
aforementioned constraints (i) and (ii) and are given
by~\cite{PruiskenBurmistrov2}
\begin{eqnarray}
    \beta_\sigma & = & \beta_\sigma^0 (\sigma_{xx}
    ,c)\hspace{0.1cm}
    - \hspace{0.15cm}\mathcal{D} (c)\sigma_{xx}^2 e^{-2\pi\sigma_{xx}} \cos 2\pi\sigma_{xy}
    \label{betainst1}\\
     \beta_\theta & = & \hspace{1.67cm}
    -\hspace{0.23cm}{\mathcal{D} (c)}\sigma_{xx}^2 e^{-2\pi\sigma_{xx}} \sin 2\pi\sigma_{xy}
    \label{betainst2}\\
    \gamma_{zc} & = & \gamma_{zc}^0 (\sigma_{xx} ,c)
    - {\mathcal{D}_\gamma (c)} \sigma_{xx} e^{-2\pi\sigma_{xx}} \cos
    2\pi\sigma_{xy} ~~~~~\label{betainst3}\\
    \beta_c &=& c(1-c) \gamma_{zc} , ~~~~~~\gamma_z =
    c \gamma_{zc} \label{betainst4}
\end{eqnarray}
Here, $\beta_\sigma^0$, $\gamma_{zc}^0$ are given by
Eqs~\eqref{beta1},~\eqref{beta2} with $\epsilon =0$ and
$\mathcal{D} (c)$, $\mathcal{D}_{\gamma} (c)$ for $c=0,1$ are
listed in Table~\ref{TZM1}. As illustrated in Fig.~\ref{RGall},
these final results fundamentally reconcile the quasi metallic
behavior (Eqs.~\eqref{beta1} -\eqref{beta4}) of the electron gas
at short distances and the quantum Hall effect
(Eq.~\eqref{Landaugap1}) that generally appears at much larger
length scales only.

In summary, the spin polarized electron gas in $D=2$ displays all
the super universal features of a $\theta$ vacuum in
asymptotically free field theory that have not been recognized
otherwise.~\cite{PruiskenBurmistrov1,PruiskenBurmistrov2}
Analogous to the theory in $D=2+\epsilon$ there are two different
critical fixed points $\sigma_{xy}^* = k+1/2$ and finite
$\sigma_{xx}^*$ in Fig.~\ref{RGall}. The plateau transitions in
the QHR therefore fall into two different universality classes
with distinctly different exponent values~\cite{Pruisken}
\begin{eqnarray}
 \nu^{-1} ={\partial \beta_\theta^*}/{\partial \sigma_{xy}} ,\,p =
 {2}/{(2+\gamma^*_{zc})} ,\, y_\sigma =
{\partial \beta_\sigma^*}/{\partial \sigma_{xx}}.\,{}
\label{exponentvalues}
\end{eqnarray}
Eqs.~\eqref{betainst1}-\eqref{betainst4}, when evaluated at the
Fermi liquid fixed point $c^* =0$, are in remarkable agreement
with the exponent values known from numerical work and the best
estimates for Eqs.~\eqref{exponentvalues} lie in the range $\nu =
2.30 - 2.38$, $p=1.22 - 1.48$ and $y_\sigma = -(0.34 -
0.42)$.~\cite{PruiskenBurmistrov1} Similar to Eq.~\eqref{Scaling}
we conclude that $\sigma_{\mu\nu}^\prime = \sigma_{\mu\nu}^\prime
(X,Y,Z)$ for finite range potentials $0<c\ll 1$ where
\begin{eqnarray}
 X=(z c T)^{-\kappa} \Delta\theta ,\qquad \Delta\theta \approx
 \sigma_{xy} - k-1/2
\end{eqnarray}
denotes the {\em relevant} scaling variable and $Y = (zc
T)^{\mu_\sigma} \Delta\sigma$, $Z = (zc T)^{\mu_c} c$
with $\Delta\sigma \approx (\sigma_{xx} -\sigma_{xx}^*)/\sigma_{xx}^*+c$ are the
{\em irrelevant} ones. The Fermi liquid exponents $\kappa =p/2\nu
= 0.29 \pm 0.04$, $\mu_\sigma =-p y_\sigma /2 = 0.26 \pm 0.05$ and
$\mu_c = p-1 = 0.35 \pm 0.15$ are clearly in conflict with the
experimental scaling results $\sigma_{\mu\nu}^\prime =
\sigma_{\mu\nu}^\prime ((zT)^{-\kappa}(\sigma_{xy} - k-1/2),(zT)^{-\mu_\sigma}(\sigma_{xx} -\sigma_{xx}^*))$~\cite{Amsterdam} with the reported
best exponent values $\kappa = 0.42 \pm 0.01 $~\cite{W.} and
$\mu_\sigma =2.5 \pm 0.5$~\cite{Amsterdam} respectively. The
experiment should in general be associated with a novel $\mathcal
F$-invariant universality class $c^* =1$ in Fig.~\ref{RGall} which
therefore disqualifies any approach based on Fermi liquid
ideas.~\cite{W.} Although the non-Fermi liquid exponent values
(Eq.~\eqref{exponentvalues}) are beyond the present theory, the
scaling results reported in this Letter nevertheless elucidate the
fundamental super universal features of the QHR which cannot be
obtained in any different
manner.~\cite{PruiskenBurmistrov2,UnifyII}

\noindent{This} research was funded in part by the Dutch Science
Foundations {\em FOM} and {\em NWO}, the {Russian Science Support
Foundation}, and the {Russian Ministry of Education and Science}.

\end{document}